\documentclass[aps,prd,preprint,preprintnumbers,nofootinbib,superscriptaddress]{revtex4-2}

\usepackage{ae}
\usepackage{bm}
\usepackage{color}
\usepackage{amsmath}
\usepackage{amssymb}
\usepackage{amsfonts}
\usepackage{graphicx}
\usepackage{slashed}
\usepackage{setspace}
\usepackage{stackrel}
\usepackage{ulem}
\usepackage{color}
\usepackage{braket}
\usepackage{siunitx}
\usepackage{float}
\usepackage{hyperref}
\usepackage{multirow}

\allowdisplaybreaks

\begin{document}

	\setlength{\unitlength}{1mm}
	\title{X-ray vacuum diffraction at finite spatio-temporal offset}
	\author{Felix Karbstein}\email{felix.karbstein@uni-jena.de}
	\affiliation{Helmholtz-Institut Jena, Fr\"obelstieg 3, 07743 Jena, Germany}
	\affiliation{GSI Helmholtzzentrum f\"ur Schwerionenforschung, Planckstra\ss e 1, 64291 Darmstadt}
	\affiliation{Theoretisch-Physikalisches Institut, Abbe Center of Photonics, \\ Friedrich-Schiller-Universit\"at Jena, Max-Wien-Platz 1, 07743 Jena, Germany}
	\author{Ricardo R. Q. P. T. Oude Weernink}\email{ricardo.reginald.quincy.philip.oude.weernink@uni-jena.de}
	\affiliation{Helmholtz-Institut Jena, Fr\"obelstieg 3, 07743 Jena, Germany}
	\affiliation{GSI Helmholtzzentrum f\"ur Schwerionenforschung, Planckstra\ss e 1, 64291 Darmstadt}
	\affiliation{Theoretisch-Physikalisches Institut, Abbe Center of Photonics, \\ Friedrich-Schiller-Universit\"at Jena, Max-Wien-Platz 1, 07743 Jena, Germany}
	\date{\today}

	\begin{abstract}
		We study the nonlinear QED signature of x-ray vacuum diffraction in the head-on collision of optical high-intensity and x-ray free-electron laser pulses at finite spatio-temporal offsets between the laser foci.
		The high-intensity laser driven scattering of signal photons outside the forward cone of the x-ray probe constitutes a prospective experimental signature of quantum vacuum nonlinearity.
		Resorting to a simplified phenomenological ad-hoc model, it was recently argued that the angular distribution of the signal in the far-field is sensitive to the wavefront curvature of the probe beam in the interaction region with the high-intensity pump.
		In this work, we model both the pump and probe fields as pulsed paraxial Gaussian beams and reanalyze this effect from first principles. We focus on vacuum diffraction both as an individual signature of quantum vacuum nonlinearity and as a potential means to improve the signal-to-background-separation in vacuum birefringence experiments.
	\end{abstract}
	
	\maketitle
	\newpage	
	\section{Introduction} \label{sec:intro}
	
\begin{figure}[h]
		\includegraphics[width=0.28\textwidth]{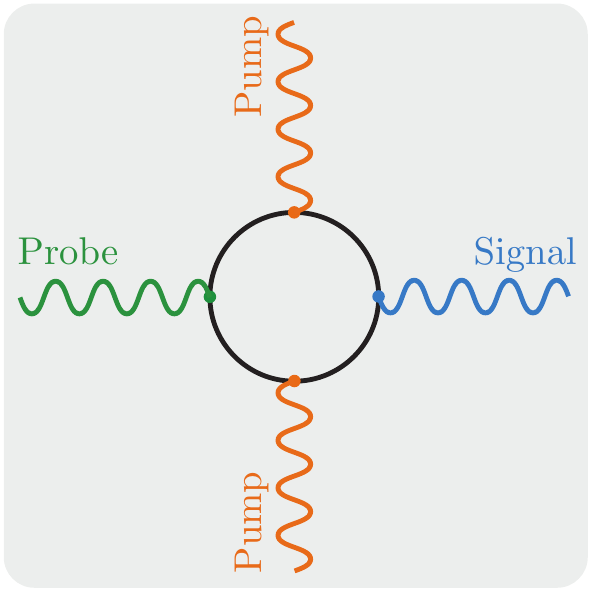}
		\caption{\label{fig:FeynGraph} Leading-order Feynman diagram giving rise to x-ray vacuum diffraction. 
			The x-ray probe interacts with the high-intensity pump via a virtual electron-positron fluctuation (solid black line). This results in signal photons differing in key properties, such as propagation direction, from the incident probe field.}
	\end{figure}

	The fluctuation of virtual particles supplements classical Maxwell theory in vacuo with effective non-linear couplings of electromagnetic fields. Within quantum electrodynamics (QED), the leading nonlinear interaction couples four electromagnetic fields; cf. Fig.~\ref{fig:FeynGraph}.
	
	The {\it critical} electric and magnetic fields $E_{\rm cr}=m_e^2 c^3/(e \hbar) \simeq 1.3 \times 10^{18}\,\si{\volt}/\si{\metre}$ and $B_{\rm cr}=E_{\rm cr}/c\simeq4\times10^9\,{\si{\tesla}}$, with electron mass $m_e\simeq511\,{\rm keV}$, serve as reference scales for the applied electromagnetic fields.
	Focusing on slowly varying fields characterized by typical frequencies much smaller than $m_e$, quantum vacuum nonlinearities can be reliably studied on the basis of the leading contribution to the Heisenberg-Euler effective Lagrangian ${\cal L}_{\rm HE}$ \cite{HeisEul,EulKock,Weisskopf:1936kd,Schw,Karplus:1950zza}. See also the pertinent reviews \cite{DittGies,Marklund,Marklund:2006my,Dunne,Piazza,King,Karbstein:2019oej} and references therein.
	In the Heaviside-Lorentz system with units where $c=\hbar=\varepsilon_0 =1$, we thus have ${\cal L}_{\rm HE}={\cal L}_{\rm MW}+{\cal L}_{\rm int}$, with classical Maxwell Lagrangian ${\cal L}_{\rm MW}=-\frac{1}{4}F_{\mu\nu}F^{\mu\nu}$ and
	\begin{align}
		\mathcal{L}_{\text{int}} \simeq \frac {2 \alpha^2}{45 m_e^4} \left[\left(F^{\mu\nu}F_{\mu\nu}\right)^2 + \frac 74 \left({}^\star\!F_{\mu\nu}F^{\mu\nu}\right)^2 \right] \,.
		\label{eq:Lint}
	\end{align}
	Here, ${}^\star\!F_{\mu\nu}$ denotes the dual field strength tensor.
	Higher-order corrections are parametrically suppressed with powers of the fine structure constant $\alpha=e^2/(4\pi)\simeq1/137$ and inverse powers of $E_{\rm cr}$ or $B_{\rm cr}$. Throughout this work we adopt the metric $g_{\mu\nu} = \text{diag}(-,+,+,+)$.
	
	Typical scenarios aiming at inducing a measurable signal of quantum vacuum nonlinearity in experiment rely on a pump-probe-type scheme: a strong pump field imprints certain properties on the quantum vacuum, which are to be probed by an additional field. At leading order the corresponding scattering amplitude is linear in the probe and quadratic in the pump field. X-ray probe photons diffracted by a strong optical high-intensity laser pump constitute a prospective quantum vacuum signature \cite{DiPiazza:2006pr, King:2010kvw,King:2010nka,Tommasini:2010fb,Inada:2017lop,Karbstein:2019bhp}. Vacuum diffraction is generic to probe photons of arbitrary polarization. Besides, it provides an additional means to improve the signal-to-background separation~\cite{Karbstein:2015xra,FelCha,Ahmadiniaz:2020kpl,Sangal:2021qeg,Mosman:2021vua} in high-intensity laser driven vacuum birefringence experiments \cite{Aleksandrov:1985,Heinzl:2006xc,Ferrando:2007pgk,Dinu:2013gaa,Dinu:2014tsa,Schlenvoigt:2016,Ataman:2018ucl,Robertson:2020nnc,Ahmadiniaz:2020lbg}. Vacuum birefringence is already actively searched for in experiments employing continuous wave lasers in combination with high-finesse cavities as probe and macroscopic magnetic fields of a few Tesla to induce the  effect \cite{Ejlli:2020yhk,Agil:2021fiq,Fan:2017fnd}. See Refs.~\cite{dEnterria:2013zqi,ATLAS:2017fur,CMS:2018erd,ATLAS:2019azn} for recent experimental evidences of light-by-light scattering in the ATLAS and CMS experiments at CERN, and Refs.~\cite{Mignani:2016fwz,Capparelli:2017mlv} for indications of the relevance of vacuum birefringence in explaining the linear polarization observed in the light from a neutron star.
	
	Microscopically, this process arises from the quasi-elastic scattering of probe photons off the strongly localized, inhomogeneous pump field. This results in signal photons featuring slightly different propagation directions than the photons comprising the incident x-ray probe. In setups based on the collision of two laser fields, the signal is maximized for exactly counter-propagating fields. The main challenge in experiment is the separation of the typically small signal from the large background of probe photons traversing the interaction region with the pump field without changing their properties. While signal photons induced in the forward cone of the probe beam are hardly discerned, those with sufficiently different emission characteristics may be. Strategies aiming at a clear directional as well as spectral separation typically require the use of more than two laser fields, see, e.g., Refs.~\cite{Mckenna:1963,Varfolomeev:1966,Rozanov:1996,Moulin:2002ya,Lundstrom:2005za,Lundin:2006wu,Bernard:2010dx,Fillion-Gourdeau:2014uua,Gies:2017ezf,King:2018wtn,Aboushelbaya:2019ncg,Klar1}.
	
	In the present work, we study x-ray vacuum diffraction at large offsets between the foci of the pump and probe beams \cite{Seino}. To this end, we perform a self-consistent first-principle-calculation not involving {\it ad hoc} assumptions. We analyze vacuum diffraction of x-ray photons both as an individual signature of quantum vacuum nonlinearity, and as an additional means to achieve a good signal-to-background separation in vacuum birefringence experiments.
	Our article is structured as follows: after briefly recalling the formalism underlying the present study in Sec.~\ref{sec:formalism}, we detail the considered laser collision scenario in Sec.~\ref{sec:stateoftheart}. Subsequently, we discuss the obtained results in Sec.~\ref{sec:results} and end with conclusions in Sec.~\ref{sec:conclusions}.

	\section{Formalism}\label{sec:formalism}
	
	Here we are interested in effects imprinted on an x-ray probe beam traversing the strong-field region generated in the focal spot of a high-intensity laser beam. The signal is assumed to be measured in the far field. In line with this, we only consider effective interaction processes which are linear in the probe on the amplitude level. The corresponding zero-to-single signal photon amplitude reads \cite{Galtsov:1971xm,Karbstein:2014fva}
	\begin{align}
		S_{(p)}({\bf k}) &= \braket{\gamma_{p}({\bf k})|\int {\rm d}^4x \,f^{\mu\nu}(x) \frac {\partial \mathcal{L}_{\text{int}}}{\partial F^{\mu\nu}}(x)|0} ,
		\label{eq:Sp}
	\end{align}
	where $\ket{0}$ denotes the vacuum state with zero signal photons and $\bra{\gamma_{p}({\bf k})}$ a state containing a single on-shell signal photon of wave vector ${\bf k}={\rm k}\hat{\bf k}$ and polarization $p$; its canonically quantized field strength tensor is $f^{\mu\nu}$; $\hat{\bf k}=(\cos\varphi\sin\vartheta,\sin\varphi\sin\vartheta,\cos\vartheta)$.
	The associated differential number of induced signal photons is obtained from this matrix element by Fermi's golden rule,
	\begin{align}
		{\rm d}^3N_p({\bf k}) = \frac {{\rm d}^3k}{(2\pi)^3} \left| S_{(p)}({\bf k})\right|^2 \,.
	\end{align}
	The total number of signal photons attainable in a polarization-insensitive measurement follows upon summation over the two transverse signal-photon polarizations.
	Upon insertion of Eq.~\eqref{eq:Lint} into Eq.~\eqref{eq:Sp} the signal photon amplitude can be represented as
	\begin{align}
		S_{(p)}({\bf k}) \simeq \frac{{\rm i}k^\mu\epsilon^{*\nu}_{(p)}({\bf k})}{\sqrt{2|{\bf k}|}}\frac {16\alpha^2}{45 m_e^4} \int {\rm d}^4x\,{\rm e}^{{\rm i}kx}\left[ F_{\rho\sigma}(x) F^{\rho\sigma}(x) F_{\mu\nu}(x)+ \frac{7}{4} {}^\star\!F_{\rho\sigma}(x) F^{\rho\sigma}(x){}^\star\!F_{\mu\nu}(x)\right]\,.
	\end{align}
	Here $\epsilon_{(p)}^\mu({\bf k})$ is the polarization vector of the signal photon, $^*$ denotes complex conjugation, and the restriction to the contribution linear in the probe field as well as $k^0=|{\bf k}|$ are implicitly assumed.
	For kinematic reasons, the signal photons are predominantly emitted into the forward cone of the probe  beam. A polarization-flipped signal-component  generically changes the polarization of the outgoing probe beam.

	\section{Considered scenario}\label{sec:stateoftheart}
	
\begin{figure}[h]
		\includegraphics[width=0.55\textwidth]{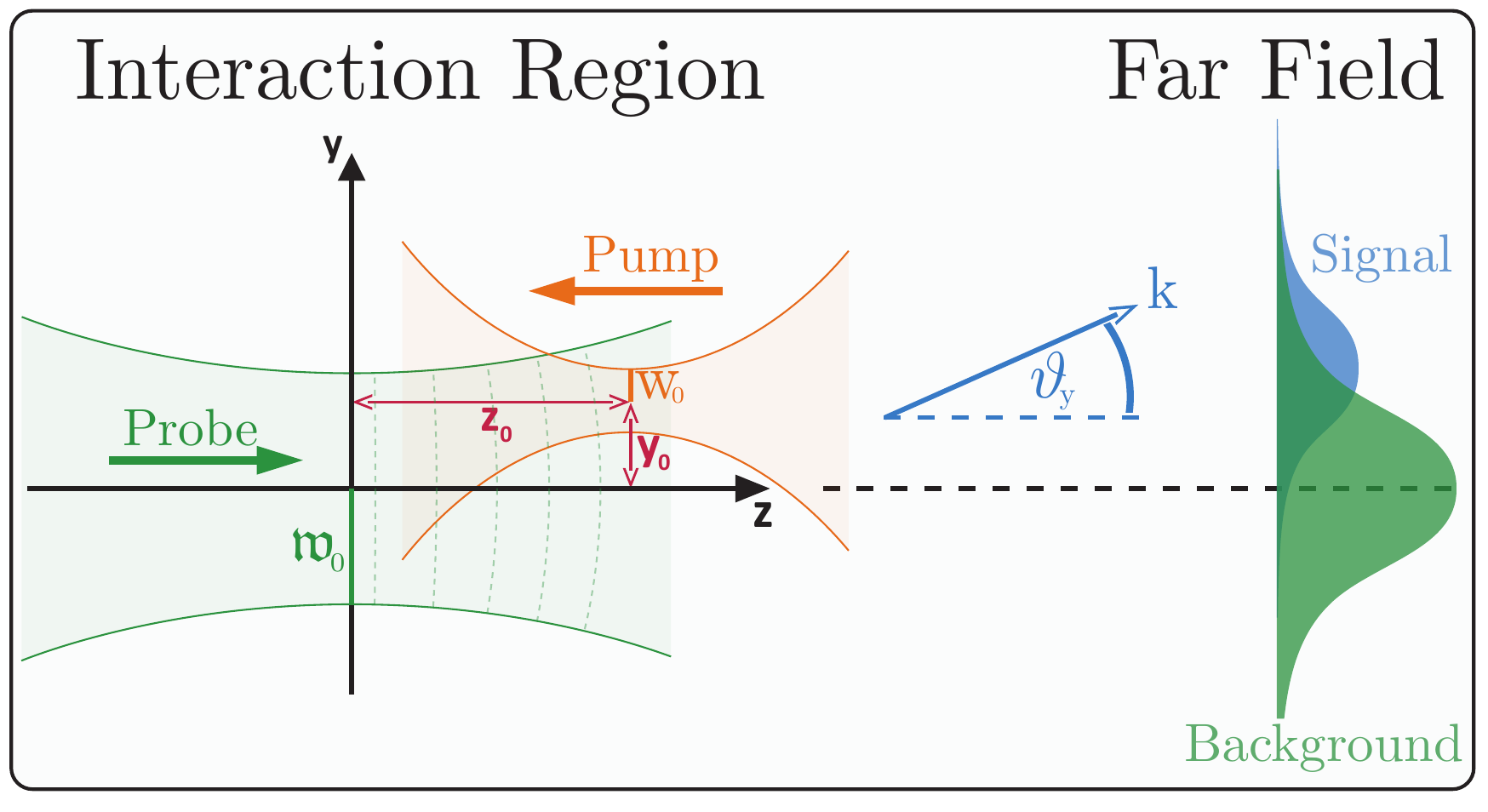}
		\caption{\label{fig:SetupSketch} Sketch of the considered scenario in the yz plane: we envision the collision of two counter-propagating laser pulses allowing for a finite offset between the beam foci. This provides a means to shift the main emission direction of the signal photons in the far field away from the forward beam axis of the probe \cite{Seino}. The probe photons traversing the interaction region unaltered constitute the background against which the signal must be discriminated; $\vartheta_{\rm y}=\vartheta|_{\varphi=\pi/2}$.}
	\end{figure}

	We consider the collision of two counter-propagating, linearly polarized laser pulses. Both laser pulses are assumed to be well described as (zeroth-order) paraxial Gaussian beams supplemented with a finite Gaussian temporal pulse envelope. The probe (pump) beam of oscillation frequency $\omega$ ($\Omega$) in the x-ray (optical) regime is propagating along the positive (negative) $\rm z$ axis. We furthermore allow for a finite spatio-temporal offset $x_0^{\mu} = (t_0,{\rm x}_0,{\rm y}_0,{\rm z}_0)$ between the foci of these beams.
	Without loss of generality the probe is assumed to be focused at ${\bf x}=0$. In accordance with the predictions of Ref.~\cite{Seino} we expect combined longitudinal and transverse focal offsets to result in an angular shift of the signal photon distribution in the far field away from the forward beam axis of the probe; see Fig.~\ref{fig:SetupSketch}.
	
	The electric field profiles of the probe and pump laser fields are denoted by ${\mathfrak E}(x)$ and $\mathcal{E}(x)$, respectively. Because inelastic scattering processes characterized by the absorption or emission of two laser photons are exponentially suppressed in comparison to quasi-elastic scattering processes \cite{Karbstein:2015xra}, in the present scenario the dominant signal turns out to be independent of the frequency of the high-intensity laser pulse. It depends only on the cycle-averaged square of the pump field profile \cite{FelCha}.
	In turn, the signal photons are predominantly emitted at frequencies close to the probe frequency $\omega$ and fulfill $\vartheta\ll1$. Correspondingly, we have \cite{Karbstein:2019oej}
	\begin{align} \label{eq:d3N}
		\left\{\begin{array}{c}
			{\rm d}^3N_{\rm tot}(\bf{k}) \\ {\rm d}^3N_\perp(\bf{k})
		\end{array}\right\} &\simeq {\rm d}^3 {\rm k}\, {\rm k}\; \frac {4\alpha^4}{45^2 \pi^3}\frac{1}{m_e^8} \left\{\begin{array}{c}
			130 - 66 \cos(2\phi) \\ 9\sin^2(2\phi)
		\end{array}\right\}	
		\bigl|\mathcal{M}({\bf k})\bigr|^2 \,,
	\end{align}
	where $\phi$ measures the angle between the  polarization vectors of the two counter-propagating beams, and we made use of the definition
	\begin{align}
		\mathcal{M}({\bf k}) &= \int {\rm d}^4x \; {\rm e}^{{\rm ik}(\hat{{\bf k}}\cdot{\bf x}-t)} \,\mathfrak{E}(x) E^2(x) \,.
		\label{eq:calM}
	\end{align}
	The first line in Eq.~\eqref{eq:d3N} gives the total number of signal photons attainable in a polarization insensitive measurement $N_{\rm tot}$, and the second one the number of signal photons scattered into a perpendicularly polarized mode $N_\perp$ constituting the signature of vacuum birefringence \cite{Toll:1952,Klein:1964zza,Baier1,Baier2,BialynickaBirula:1970vy,Adler:1971wn} in a high-intensity laser experiment. Note, that $N_{\rm tot}$ is maximized for $\phi=\pi/2$ and $N_\perp$ for $\phi=\pi/4$ \cite{Karbstein:2019bhp}.
	When providing numerical results for the signal photon numbers below we will always implicitly adopt the choice of $\phi$ maximizing the considered observable.
	
	We also note that for the considered scenario involving an x-ray probe featuring photon energies in the ${\cal O}(10)\,{\rm keV}$ regime and pulse durations of ${\cal O}(10)\,{\rm fs}$, the signal photon spectrum is strongly peaked at the oscillation frequency $\omega$ of the probe and decays rapidly to zero within a fraction of an electron-volt \cite{Karbstein:2015xra}.
	Hence, when aiming at performing the integration over the signal photon energy $\rm k$ in Eq.~\eqref{eq:d3N} it serves as an excellent approximation to identify the powers of $\rm k$ in the overall prefactor multiplying the Fourier integral with $\omega$ and to extend the limits of the Gaussian integral over the entire real $\rm k$ axis \cite{Karbstein:2018omb}.
	This results in the following far-field angular distributions of the signal photons,
	\begin{align} \label{eq:d2N}
		\left\{\begin{array}{c}
			{\rm d}^2N_{\rm tot}(\varphi,\vartheta) \\ {\rm d}^2N_\perp(\varphi,\vartheta)
		\end{array}\right\} &\simeq {\rm d}\varphi\,{\rm d}\!\cos\vartheta\,\frac{4\alpha^4}{45^2 \pi^3}\frac{\omega^3}{m_e^8} \left\{\begin{array}{c}
			130 - 66 \cos(2\phi) \\ 9\sin^2(2\phi)
		\end{array}\right\}	
		\int_{-\infty}^\infty {\rm dk}\,\bigl|\mathcal{M}({\bf k})\bigr|^2 \,.
	\end{align}
	The polar angle $\vartheta$ is measured from the forward beam axis of the probe laser, and the azimuthal angle $\varphi$ parameterizes rotations about the beam axis. An angle of $\varphi=0$ ($\pi/2$) is associated with signal photon emission in the xz (yz) plane.

	As detailed above, the fields entering Eq.~\eqref{eq:calM}, are the probe field profile,
	\begin{align}
		\mathfrak{E}(x) &= \mathfrak{E}_0\, {\rm e}^{-\left(\frac{{\rm z}-t}{T/2}\right)^2} \frac {\mathfrak{w}_0}{\mathfrak{w}({\rm z})} \,{\rm e}^{-\frac {{\rm x}^2+{\rm y}^2}{\mathfrak{w}^2({\rm z})}} \cos\biggl(\omega({\rm z}-t) + \frac {\omega({\rm x}^2+{\rm y}^2)}{2R({\rm z})}-\arctan\Bigl(\frac {\rm z}{{\mathfrak z}_R} \Bigr) \biggr)\,,\label{eq:FieldAmpProbe}
	\end{align}
	and the cycle-averaged squared pump field given by
	\begin{align}
		\mathcal{E}^2(x)\simeq \frac{1}{2}\mathcal{E}_0^2\,{\rm e}^{-2\left(\frac{{\rm z}-{\rm z}_0+t-t_0}{\tau/2}\right)^2}\!\Bigl(\frac{w_0}{w({\rm z}-{\rm z}_0)}\Bigr)^2 {\rm e}^{-2\frac{({\rm x}-{\rm x}_0)^2+({\rm y}-{\rm y}_0)^2}{w^2({\rm z}-{\rm z}_0)}}\,.
		\label{eq:FieldAmpPump}
	\end{align}
	Here, $\mathfrak{z}_R$ denotes the Rayleigh range, and $\mathfrak{w}({\rm z})=\mathfrak{w}_0\sqrt{1+({\rm z}/\mathfrak{z}_R)^2}$ measures the transversal widening of the probe beam as a function of the longitudinal coordinate $\rm z$; its beam waist is $\mathfrak{w}_0$, and the radius of curvature of its wavefronts is $R({\rm z})=\frac{\mathfrak{z}_R^2}{\rm z}(\frac{\mathfrak{w}({\rm z})}{\mathfrak{w}_0})^2$. 
	Analogously, for the pump beam we have $w({\rm z})=w_0\sqrt{1+({\rm z}/{\rm z}_R)^2}$, with beam waist $w_0$ and Rayleigh range ${\rm z}_R$.
	The peak field amplitude $\mathfrak{E}_0$ ($E_0$) of the probe (pump) field can be expressed in terms of its pulse energy $\mathfrak{W}$ ($W$), waist spot size and pulse duration $T$ ($\tau$), as $\mathfrak{E}_0\simeq2(\frac{8}{\pi})^{1/4}\sqrt{\frac{\mathfrak{W}}{\pi\mathfrak{w}_0^2T}}$ and $\mathcal{E}_0\simeq2(\frac{8}{\pi})^{1/4}\sqrt{\frac{W}{\pi w_0^2\tau}}$, respectively \cite{Karbstein:2019oej}.
	
	Equation~\eqref{eq:FieldAmpProbe} only amounts to a solution of the paraxial Helmholtz equation for $\mathfrak{z}_R=\omega\mathfrak{w}_0^2/(2M^2)$ with $M^2\equiv1$. On the other hand, a {\it beam quality factor} $M^2\neq1$ is widely employed to phenomenologically capture important characteristics of the field profiles of experimentally realistic laser beams deviating from ideal fundamental Gaussian beams \cite{Siegman:1993,SalehTeich:2019}. 
	Particularly as the {\it prototype experiment} discussed by Ref.~\cite{Seino} invokes a choice of $M^2=10$, in the present study we account for a generic factor of $M^2$. Furthermore, we note that by construction the approximate beam profiles~\eqref{eq:FieldAmpProbe} and \eqref{eq:FieldAmpPump} are only expected to allow for a reliable description of the colliding laser pulses as long as the radial far-field divergence of the probe (pump) beam $\theta\simeq\frac{\mathfrak{w}_0}{\mathfrak{z}_R}$ ($\Theta\simeq\frac{w_0}{{\rm z}_R}$) fulfills $\theta\ll1$ ($\Theta\ll1$). At the same time the pulse durations should be sufficiently long, such that the condition $\{\omega T,\Omega\tau\}\gg1$ holds; cf., e.g., Ref.~\cite{King:2012aw}.
	
	Upon plugging Eqs.~\eqref{eq:FieldAmpProbe} and \eqref{eq:FieldAmpPump} into Eq.~\eqref{eq:calM}, all integrations apart from the one over the longitudinal coordinate $\rm z$ can be performed analytically. This is similar to the head-on collision scenario analyzed in Ref.~\cite{FelCha} involving an infinite Rayleigh range approximation for the probe beam: when the widening of at least one of the driving laser beams as a function of its longitudinal coordinate is explicitly taken into account, the integration over this coordinate can no longer be performed analytically in any simple way. Due to the fact that in the present case we consistently account for the widening of both beams, the situation is even more complicated.
	
	Rewriting the modulus square of the amplitude~\eqref{eq:calM} as $|{\cal M}({\bf k})|^2= {\cal M}({\bf k}) {\cal M}^*({\bf k})$ and denoting the variables of the two independent longitudinal integrations as $\rm z$ and $\bar{\rm z}$, respectively, the integration over the signal photon energy $\rm k$ in Eq.~\eqref{eq:d2N} can be readily carried out analytically. This results in
	\begin{align} \label{eq:intMmod2}
		\int_{-\infty}^\infty {\rm dk}\,\bigl|\mathcal{M}({\bf k})\bigr|^2 &=\frac {16}{\pi}\frac{ \mathfrak{W} W^2}{2 T^2 + \tau^2}\, {\rm e}^{- \frac{1}{8}\frac {\tau^2}{2T^2 + \tau^2}(T\omega)^2} 
		\nonumber \\
		&\quad \times \frac{1}{\mathfrak{w}_0^2}\int_{-\infty}^{\infty}{\rm dz} \int_{-\infty}^{\infty}{\rm d}\bar{\rm z} \;\; \frac {G({\rm z}) G^*(\bar{\rm z})}{\sqrt{H({\rm z},\bar{\rm z},\vartheta)}}\, {\rm e}^{2\frac{\left[ h({\rm z},\varphi,\vartheta) + h^*(\bar{\rm z},\varphi,\vartheta) \right]^2}{H({\rm z},\bar{\rm z},\vartheta)}} \,,
	\end{align}
	where we introduced the auxiliary functions
	\begin{subequations} \label{eq:Abbrev}
		\begin{align}
			G(\rm z) &= \frac{g({\rm z})}{1-{\rm i}\frac{\rm z}{{\mathfrak z}_R}}\,	{\rm e}^{- 4\frac {2  (2{\rm z}-{\rm z}_0-t_0)^2 + {\rm i} (T\omega) T{\rm z} }{2T^2 + \tau^2}  - 2\frac{[1-g({\rm z})]({\rm x}_0^2 + {\rm y}_0^2)}{w^2(\rm z-{\rm z}_0)}} \,, \label{eq:Abbrev:G} \\
			g({\rm z})&= \frac{2}{2+(1+{\rm i}M^2\frac{{\rm z}}{\mathfrak{z}_R})(\frac{w({\rm z}-{\rm z}_0)}{\mathfrak{w}({\rm z})})^2}\,, \\
			h({\rm z},\varphi,\vartheta) &= \frac{1}{8}\frac{\tau^2}{2 T^2 + \tau^2}T\omega + {\rm i}\Bigl(\cos\vartheta + \frac{2T^2 - \tau^2}{2T^2 + \tau^2}\Bigr)\frac{\rm z}{T} + {\rm i}g({\rm z})\frac{{\bf x}_0\cdot\hat{\bf k}|_{\vartheta=\frac{\pi}{2}}}{T} \sin\vartheta \,, \\
			H({\rm z},\bar{\rm z},\vartheta) &= \frac{\tau^2}{2T^2+\tau^2} + \frac{g({\rm z})w^2({\rm z}-{\rm z}_0) + g^*(\bar{\rm z})w^2(\bar{\rm z}-{\rm z}_0) }{T^2}\sin^2\vartheta \,.
		\end{align}
	\end{subequations}
	The dependence of Eq.~\eqref{eq:intMmod2}  on the azimuthal angle $\varphi$ is encoded in the function $h({\rm z},\varphi,\vartheta) $ via ${\bf x}_0\cdot\hat{\bf k}|_{\vartheta=\frac{\pi}{2}}={\rm x}_0\cos\varphi+{\rm y}_0\sin\varphi$.
	
	Equations.~\eqref{eq:FieldAmpProbe} and \eqref{eq:FieldAmpPump} imply that the choice of $t_0=-{\rm z}_0$ ensures that the temporal envelopes of the probe and pump pulses reach their maxima simultaneously at the longitudinal focus coordinate of the latter. As this choice typically maximizes the signal, we adopt it in the remainder of this article.
	
	When employing an {\it infinite Rayleigh range approximation} for both beams, characterized by formally taking the limit $\{\mathfrak{z}_R,{\rm z}_R\}\to\infty$ while promoting the beam waists to constant values as $w({\rm z}-{\rm z}_0)\to w_0$ and $\mathfrak{w}({\rm z})\to\mathfrak{w}({\rm z}_0)$, the remaining integrations in Eq.~\eqref{eq:intMmod2} become Gaussian and can be performed analytically.
	In this specific limit, finite offsets between the foci of the probe and pump laser beams only result in a damping of the signal relatively to a collision at zero offset. The identification of the waist sizes with their values at ${\rm z}={\rm z}_0$ is motivated by the fact that the signal photons are predominantly originating from the space-time region where the pump field strength peaks: for $t_0=-{\rm z}_0$, this maximum is reached at ${\rm z}={\rm z}_0$. For the laser pulse collision scenario considered here, the infinite Rayleigh range approximation should allow for reliable results given that $\{T,\tau\}\ll\{\mathfrak{z}_R,{\rm z}_R\}$ and $z_0\ll\mathfrak{z}_R$. However, by construction it accounts neither for beam widening effects coming along with a reduction of the peak field with increasing distance from the focus, nor for beam curvature effects.
	The former point implies that especially for large pulse durations the strong-field region -- and thus the signal -- tends to be overestimated.
	
	Making use of the fact that signal photon emission predominantly occurs in the vicinity of ${\rm z}={\rm z}_0$, we can infer some qualitative analytical scalings describing the behavior of the directional emission characteristics of the signal in dependence of the focal offset $\vec{x}_0$.
	To this end, we identify ${\rm z}=\bar{\rm z}={\rm z}_0$ in the integrand of Eq.~\eqref{eq:intMmod2} and expand the terms in the exponential up to quadratic order in $\vartheta\ll1$.
	In general, the exponential dependence of Eq.~\eqref{eq:intMmod2} on $\vartheta$ can be cast into the form $\exp\{-2(\vartheta-\Delta\vartheta_{\rm s})^2/\theta_{\rm s}^2\}$, where
	\begin{equation}
		\theta_{\rm s}(\varphi)\simeq\frac{\frac{2}{\omega\mathfrak{w}({\rm z}_0)}\bigl\{[1+2(\frac{\mathfrak{w}({\rm z}_0)}{w_0})^2]^2+(\frac{{\rm z}_0M^2}{\mathfrak{z}_{\rm R}})^2\bigr\}}{\sqrt{[1+2(\frac{\mathfrak{w}({\rm z}_0)}{w_0})^2]\bigl\{[1+2(\frac{\mathfrak{w}({\rm z}_0)}{w_0})^2]^2+(\frac{{\rm z}_0M^2}{\mathfrak{z}_{\rm R}})^2\bigr\}-\frac{2T^2+\tau^2}{\tau^2}(\frac{8}{\omega T}\frac{\mathfrak{w}({\rm z}_0)}{w_0}\frac{{\rm z}_0M^2}{\mathfrak{z}_{\rm R}}\frac{{\bf x}_0\cdot\hat{\bf k}}{w_0}\big|_{\vartheta=\frac{\pi}{2}})^2}}, \label{eq:thetas}
	\end{equation}
	can be identified with the radial signal divergence, and
	\begin{equation} 
		\Delta\vartheta_{\rm s}(\varphi)\simeq\frac{\frac{{\rm z}_0M^2}{\mathfrak{z}_R}(\frac{\mathfrak{w}({\rm z}_0)}{w_0})^2\,\theta_{\rm s}^2}{[1+2(\frac{\mathfrak{w}({\rm z}_0)}{w_0})^2]^2+(\frac{{\rm z}_0M^2}{\mathfrak{z}_{\rm R}})^2}\, \omega\,{\bf x}_0\cdot\hat{\bf k}|_{\vartheta=\frac{\pi}{2}} \,, \label{eq:deltathetas}
	\end{equation}
	with an angular shift of the signal emission direction away from the forward beam axis of the probe.
	For vanishing transverse offset, ${\rm x}_0={\rm y}_0=0$, Eq.~\eqref{eq:thetas} is independent of both the pump and probe pulse durations, and Eq.~\eqref{eq:deltathetas} vanishes identically. 
	On the other hand, when adopting an infinite Rayleigh range approximation for the probe beam Eq.~\eqref{eq:thetas} reduces to the result for the signal divergence derived in Ref.~\cite{Karbstein:2018omb}.
	
	For $\omega T\gg1$, as considered here, the variation of the estimated signal divergence~\eqref{eq:thetas} with the transverse offset -- and thus also the polar angle $\varphi$ -- should be very mild.
	Completely neglecting the subleading dependence on the transverse coordinate, Eqs.~\eqref{eq:thetas} and \eqref{eq:deltathetas} simplify substantially and read
	\begin{equation}
		\theta_{\rm s}\simeq \frac{2}{\omega\mathfrak{w}({\rm z}_0)}\sqrt{\frac{[1+2(\frac{\mathfrak{w}({\rm z}_0)}{w_0})^2]^2+(\frac{{\rm z}_0M^2}{\mathfrak{z}_{\rm R}})^2}{1+2(\frac{\mathfrak{w}({\rm z}_0)}{w_0})^2}}\,,\label{eq:thetas2}
	\end{equation}
	and
	\begin{equation} 
		\Delta\vartheta_{\rm s}(\varphi)\simeq \frac{4M^2}{\omega\mathfrak{z}_R}\,\frac{{\rm z}_0({\rm x}_0\cos\varphi+{\rm y}_0\sin\varphi)}{2 \mathfrak{w}({\rm z}_0)^2 + w_0^2} \,. \label{eq:deltathetas2}
	\end{equation}
	Equation~\eqref{eq:deltathetas2} indicates that, as to be expected from the beam curvature of the probe in the interaction region with the pump, for ${\rm z}_0>0$ the main emission direction is shifted in the direction of the transverse offset.
	
	Finally, we note that the far-field angular decay of the number of probe photons $\mathfrak{N}\simeq\mathfrak{W}/\omega$, which traverse the interaction region with the pump essentially unmodified, can be expressed as
	\begin{equation}
		{\rm d}^2\mathfrak{N}\simeq{\rm d}\varphi\,{\rm d}\!\cos\vartheta\,\frac{4\mathfrak{N}}{2\pi}\Bigl(\frac{\mathfrak{z}_R}{\mathfrak{w}_0}\Bigr)^2\,{\rm e}^{-2(\frac{\mathfrak{z}_R}{\mathfrak{w}_0})^2\vartheta^2}\,. \label{eq:d2Nprobe}
	\end{equation}
	This is an important quantity, as it constitutes the background against which the signal photons have to be discriminated in experiment. Equation~\eqref{eq:d2Nprobe} implies that the $1/{\rm e}^2$ far-field radial divergence of the probe is given by $\theta=\mathfrak{w}_0/\mathfrak{z}_R=2M^2/(\omega\mathfrak{w}_0)$.
	
	\section{Results}\label{sec:results}
	
	Subsequently, we study the nonlinear QED signatures of x-ray photon diffraction and vacuum birefringence in laser pulse collisions at large focal offsets for experimentally realistic parameters.
	In a first step, we aim at verifying the impact of probe beam curvature effects on the far-field angular distribution of the signal photons put forward by Ref.~\cite{Seino} on the basis of a simplified phenomenological {\it ad hoc} model.
	The model of Ref.~\cite{Seino} bases on an infinite Rayleigh range approximation for the probe beam \cite{Karbstein:2018omb}, which is {\it a posteriori} augmented with beam curvature effects.
	In a second step, we analyze if a finite offset between the foci of the driving laser pulses may constitute a handle to enhance the signal-to-background separation in experiment.
	To this end, we consider two sets of parameters characterizing the driving laser pulses, which we refer to as Setups~A and B. For their explicit parameters, see Tab.~\ref{tab:LaserParam}.
	
	\begin{table}
		\begin{tabular}{l||c|c|}
			& \textbf{Setup A} & \textbf{Setup B} \\
			\hline \hline
			\textbf{Probe Parameters:} & & \\
			waist $\mathfrak{w}_0$ [\si{\micro\metre}] & $6$ & $3$ \\
			pulse duration $T$ [\si{\femto\second}] & $17$ & $17$, $220$ \\
			frequency $\omega$ [\si{\kilo\eV}] & $9.8$ & $12.914$ \\
			pulse energy $\mathfrak{W}$ [\si{\milli\joule}] & $0.47$ & $2.07$ \\
			beam quality factor $M^2$ & $10$ & $1\ldots10$\\
			\hline
			\textbf{Pump Parameters:} & & \\
			waist $w_0$ [\si{\micro\metre}] & $9.8$ & $1.0$ \\
			pulse duration $\tau$ [\si{\femto\second}] & $40$ & $40$ \\
			Rayleigh range ${\rm z}_R$ [\si{\micro\metre}] & $377.15$ & $3.93$ \\	
			pulse energy $W$ [\si{\joule}] & $2.1\times 10^{-4}$ & $12.5$ \\
		\end{tabular}
		\caption{\label{tab:LaserParam} Parameter sets characterizing the driving laser pulses considered in the present article. For Setup~B we consider two different probe pulse durations and various beam quality factors.}
	\end{table}
	Setup~A corresponds to the {\it prototype experiment} analyzed in detail in Ref.~\cite{Seino}. We employ it as benchmark configuration. Reference~\cite{Seino} studied two specific focal offsets: (a) a purely longitudinal offset along the z axis, and (b) the same longitudinal offset complemented with a transverse one along the y axis. See Tab.~\ref{tab:ShiftOrd} for the specific offsets considered in Ref.~\cite{Seino}. While (b) parameterizes the actual focal offset implemented in the prototype experiment performed by the authors of Ref.~\cite{Seino}, the purely longitudinal offset (a) is intended to serve as a reference allowing to highlight the impact of the finite transverse offset considered in (b) on the far-field angular distribution of the signal photons.
	\begin{table}
		\begin{tabular}{l||c|c|}
			&  \multicolumn{2}{c|}{\bf Setup A} \\
			\hline \hline
			\textbf{Focal offset:} & (a) & (b) \\
			x-shift ${\rm x}_0$ [\si{\micro\metre}] & $0$ &$0$ \\
			y-shift ${\rm y}_0$ [\si{\micro\metre}] & $0$& $3.7$ \\
			z-shift ${\rm z}_0$ [\si{\metre}] & $0.85$ &$0.85$ \\
		\end{tabular}
		\caption{\label{tab:ShiftOrd} Specific offsets between the  foci of the probe and pump laser beams considered for the benchmark scenario. Note, that the particular offsets chosen here fulfill ${\rm y}_0={\cal O}(1)\mathfrak{w}_0$ and ${\rm z}_0={\cal O}(10)\mathfrak{z}_R$.}
	\end{table}
	Due to the cylindrical symmetry of the considered scenario about the beam axis of the probe beam, we can without loss of generality set ${\rm x}_0=0$ and limit our discussion to transverse offsets ${\rm y}_0\neq0$. Besides, in accordance with Ref.~\cite{Seino} for the analysis of Setup A we limit ourselves to a probe beam quality factor of $M^2=10$.
	
	For Setup~B we adopt the parameters of a prospective discovery experiment for vacuum diffraction and birefringence, which can be implemented with state-of-the-art technology. Here, we envision larger pulse energies, a harder focusing of the driving laser pulses, and employ probe beam quality factors in the representative range $1\leq M^2\leq 10$. The latter values span the entire parameter regime from  $M^2=1$ for the ideal fundamental Gaussian beam to the fairly large value of $M^2=10$ invoked to describe the experimentally realistic laser beam employed by Ref.~\cite{Seino} in their prototype experiment. As the x-ray polarimeter required for the experimental detection of vacuum birefringence generically increases an originally available FEL pulse duration quite substantially \cite{Mosman:2021vua}, we moreover consider two distinct probe pulse durations, differing by about one order of magnitude.

	\subsection{Benchmark Scenario}\label{subsec:benchmark}
	
\begin{figure}
		\includegraphics[width=0.7\textwidth]{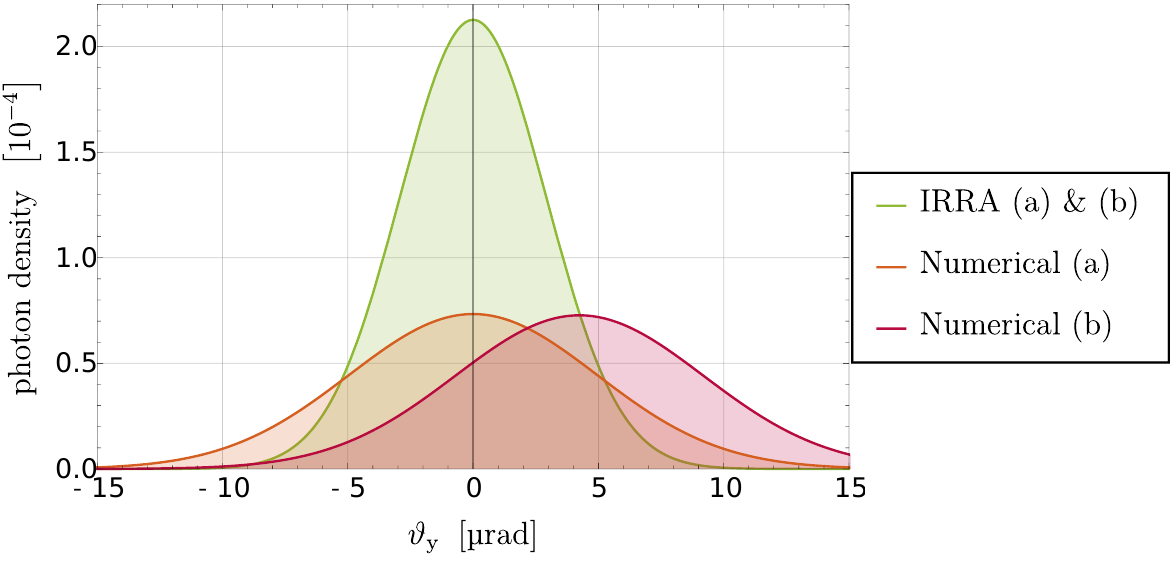}
		\caption{\label{fig:Recreation} Far-field angular distribution of the signal photons $\textrm{d}^2N_{\rm tot}(\varphi,\vartheta)/(\textrm{d}\varphi\,\textrm{d}\!\cos\vartheta)$ for Setup~A and the focal offsets (a) and (b) as a function of $\vartheta_{\rm y}$. We depict results obtained from a numerical evaluation of Eq.~\eqref{eq:d2N} and from an infinite Rayleigh range approximation (IRRA).}
	\end{figure}

	First we adopt the parameters of Setup A. Figure~\ref{fig:Recreation} shows the far-field angular distribution of the signal photons $N_{\rm tot}$ as a function of $\vartheta_{\rm y}=\vartheta|_{\varphi=\pi/2}$.
	Here, we present results obtained from a direct numerical evaluation of Eq.~\eqref{eq:d2N} for the laser pulse parameters of Setup A and the focal offsets (a) and (b).
	For comparison, we also depict the results obtained from an	infinite Rayleigh range approximation; note that in the present case we have $\mathfrak{w}({\rm z}_0)\simeq 57 \,\si{\micro\metre}$.
	The latter curves essentially fall on top of each other.
	See Tab.~\ref{tab:RecreationPara} for the parameters characterizing these results, namely the angular shifts $\Delta\vartheta_{\rm s,x}$, $\Delta\vartheta_{\rm s,y}$ of the peaks from the forward beam axis of the probe in x ($\varphi=0$) and y ($\varphi=\pi/2$) directions, the associated radial $1/{\rm e}^2$ divergences $\theta_{\rm s,x}$, $\theta_{\rm s,y}$, and the signal photon numbers $N_{\rm tot}$.
	Obviously, the maximal attainable number of signal photons follows upon integration of Eq.~\eqref{eq:d2N} over the full solid angle.
	\begin{table}
		\begin{tabular}{c|c||c|c|c|c|}
			Focal offset& Method & $\Delta\vartheta_{\rm s,y}$ [\si{\micro\radian}] & $\theta_{\rm s,x}$ [\si{\micro\radian}] & $\theta_{\rm s,y}$ [\si{\micro\radian}] & $N_{\rm tot}$ \\
			\hline\hline
			& Numerical & $0.00$ & $9.90$ & $9.90$ & $1.13\times 10^{-14}$ \\
			(a)& Eqs.~\eqref{eq:thetas} and \eqref{eq:deltathetas2} & $0.00$ & $9.93$ & $9.93$ & -- \\
			& IRRA & $0.00$ & $5.83$ & $5.83$ & $1.14\times 10^{-14}$ \\
			\hline
			& Numerical & $4.24$ & $9.90$ & $9.90$ & $1.12\times 10^{-14}$ \\
			(b)& Eqs.~\eqref{eq:thetas} and \eqref{eq:deltathetas2} & $0.00$ & $9.93$ & $9.93$ & -- \\
			& IRRA  & $0.00$ & $5.83$ & $5.83$ & $1.13\times 10^{-14}$ 
		\end{tabular}
		\caption{\label{tab:RecreationPara} Characteristic properties of the angular signal photon distribution for Setup~A and the focal offsets (a) and (b). For comparison we provide results obtained by an infinite Rayleigh range approximation (IRRA). For all parameter sets highlighted in this table we find $\Delta\vartheta_{\rm s,x}\simeq0$.}
	\end{table}
	The values for $\theta_{\rm s,x}$ and $\theta_{\rm s,y}$ are extracted from Gaussian fits to the numerical data.
	Besides, we provide the corresponding values of the analytical estimates \eqref{eq:thetas} and \eqref{eq:deltathetas2}.
	We emphasize the remarkably good agreement with the findings of Ref.~\cite{Seino} determined on the basis of a phenomenological {\it ad hoc} model.
	
	The numerical results show that for ${\rm z}_0\neq0$ a transverse focal shift ${\rm y}_0>0$ correlates with an angular shift of the maximum of the signal photon distribution in y direction. This matches the prediction of Eq.~\eqref{eq:deltathetas2}. Besides, for the small transverse offset ${\rm y}_0\ll \mathfrak{w}({\rm z}_0)$ considered here, the signal photon number for the focal offset (b) is only slightly reduced relatively to the one attainable with the purely longitudinal offset (a).
	
	As the parameters of Setup A are characterized by $\{\mathfrak{z}_R,{\rm z}_R\}\gg\{T,\tau\}$, the signal photon numbers derived by an infinite Rayleigh range approximation rather accurately reproduce the full numerical results.
	At the same time, the infinite Rayleigh approximation completely neglects wavefront curvature effects, and thus cannot reproduce the angular shift $\Delta\vartheta_{\rm s,y}$ of the signal for ${\rm y}_0>0$.
	Wavefront curvature effects are expected to be sizable for laser pulse collisions at longitudinal offset of ${\rm z}_0\gtrsim\mathfrak{z}_R$.
	Also another characteristic feature of the numerically determined signal photon distributions can be attributed to these effects. Their divergences are somewhat larger than those derived on the basis of an infinite Rayleigh approximation for both beams: the finite opening angle of the probe beam at the collision point generically tends to increase the signal divergence.
	
	Finally, we emphasize that for Setup A the background of the probe photons $\mathfrak N={\cal O}(10^{12})$ which traverse the interaction region with the high-intensity pump essentially unmodified vastly dominates the induced signal $N_{\rm tot}={\cal O}(10^{-14})$; cf. Tab.~\ref{tab:RecreationPara}. While an increase of ${\rm y}_0$ tends to increase $\Delta\vartheta_{\rm s,y}$, simultaneously the signal photon number rapidly drops as less and less probe photons see the strong pump field.

	\subsection{Prospective Discovery Experiment}
	
	Subsequently, we focus on Setup B.
	First, we analyze the dependence of different properties characterizing the far-field directional distribution of the signal photons on various input parameters.
	Second, we study the perspectives of using focal offsets to induce an angular shift of the signal photon distribution away from the forward beam axis of the probe as a potential means to improve the signal-to-background separation in experiment.
	
\begin{figure}
		\includegraphics[width=0.8\textwidth]{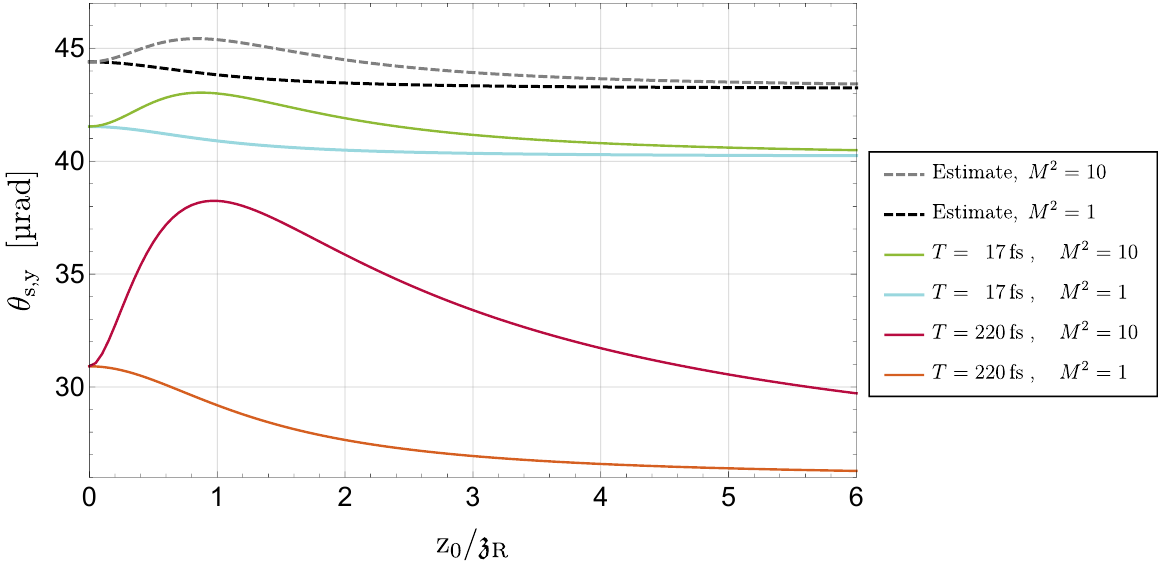}
		\caption{\label{fig:Z0Div} Signal divergence $\theta_{\rm s,y}$ for Setup B as a function of the longitudinal offset ${\rm z}_0$ measured in multiples of the probe Rayleigh range $\mathfrak{z}_R$. The legend specifies the parameters characterizing the various curves depicted here (from top to bottom).}
	\end{figure}

	Figure~\ref{fig:Z0Div} illustrates the behavior of the signal divergence $\theta_{\rm s, y}$ as function of the longitudinal offset ${\rm z}_0$ for two different probe pulse durations $T$, and two different values of the probe beam quality factor $M^2$.
	Though the analytical estimates~\eqref{eq:thetas2} for $\theta_{\rm s,y}$ differ sizably from the corresponding numerical results, they show the same general behavior on ${\rm z}_0$.
	The encountered deviations are to be expected as the estimate does not at all depend on the pump and probe pulse durations. As its derivation assumes the signal to arise from the vicinity of ${\rm z}={\rm z}_0$ only, for fixed other parameters the estimate~\eqref{eq:thetas2} should approach the corresponding exact result in the formal limit of $T\to0$.
	The behavior of the curves in Fig.~\ref{fig:Z0Div} is in line with this expectation.
	In addition, Fig.~\ref{fig:Z0Div} unveils an interesting feature:
	while the curves for $M^2=1$ decrease monotonically with increasing ${\rm z}_0$, the ones for $M^2=10$ exhibit a maximum at ${\rm z}_0>0$. Analyzing the occurrence of the sign change of  $\partial^2_{{\rm z}_0}\theta_{\rm s}\big|_{{\rm z}_0=0}$ as a function of $M^2$, with $\theta_{\rm s}$ given by Eq.~\eqref{eq:thetas2}, it can be straightforwardly inferred that such a maximum occurs for beam quality factors fulfilling $M^2\geq \sqrt{1+2( \frac {\mathfrak{w}_0}{w_0} )^2}$. For the parameters of Setup~B considered here, this criterion becomes $M^2\gtrsim4.36$.
	However, we emphasize here that the formation of a non-trivial maximum of the signal photon divergence at ${\rm z}_0>0$ may likely be an artifact of the probe field with $M^2\neq1$ ceasing to be a self-consistent solution of the paraxial Helmholtz equation. Aiming at resolving if this maximum amounts to a physical effect or not, one would need to compare the present results with those based upon a more advanced beam model for the pump solving the (paraxial) Helmholtz equation. In fact, we are not aware of any physical reason hinting at the formation of such a feature. Therefore, we subsequently only discuss explicit results for the theoretically solid and self-consistent case characterized by $M^2=1$; see also Eq.~\eqref{eq:thetasbytheta} below which anyhow suggests that the best signal-to-background separation should be attainable with the smallest possible value of $M^2\geq1$.
	
\begin{figure}
		\includegraphics[width=1\textwidth]{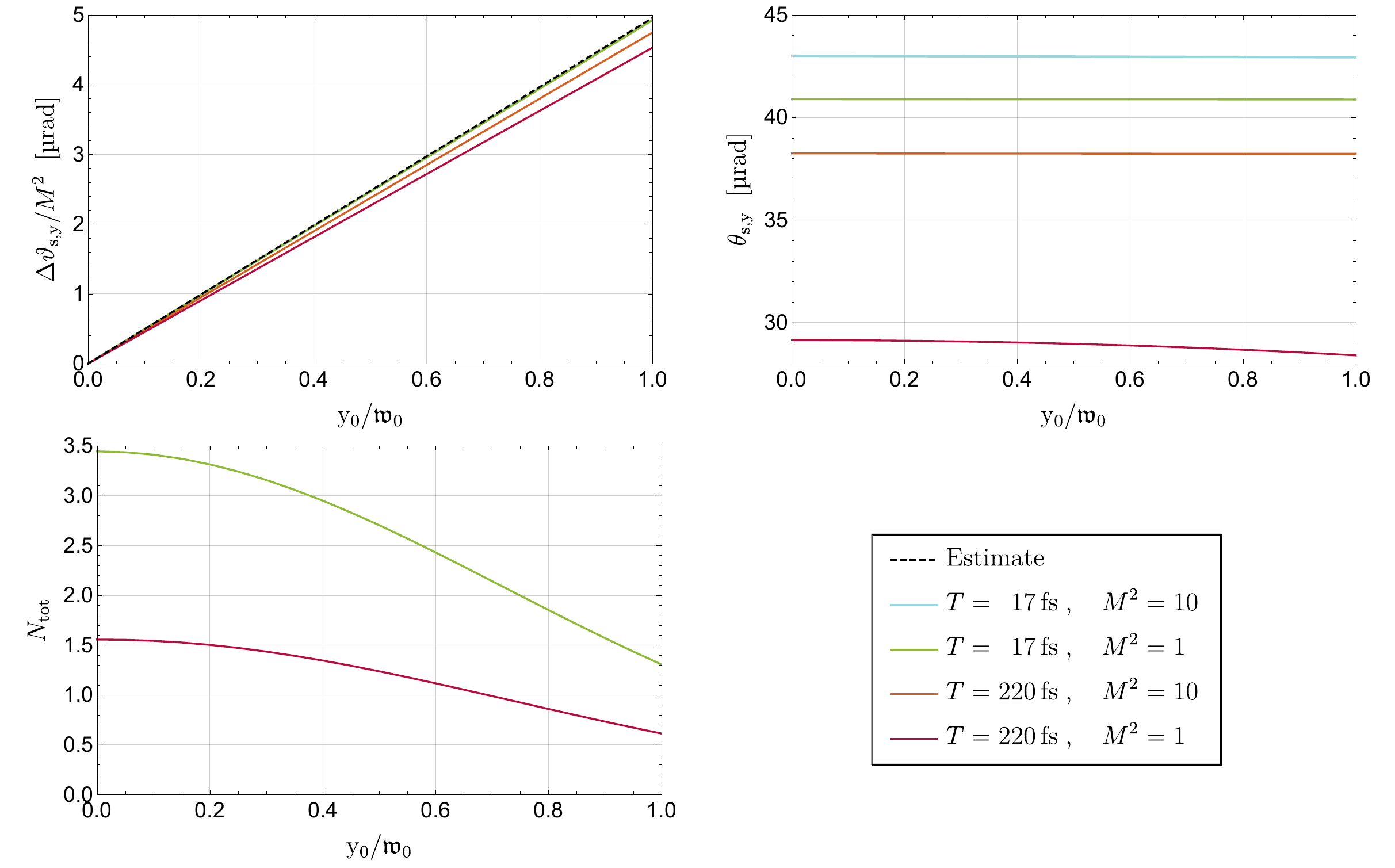}
		\caption{\label{fig:Z0Y0} Dependence of the parameters characterizing the signal for Setup B (angular shift $\Delta\vartheta_{\rm s,y}$, radial divergence $\theta_{\rm s,y}$, and signal-photon yield $N_{\rm tot}$) as a function of the transverse offset ${\rm y_0}$ measured in units of the probe waist $\mathfrak{w}_0$; ${\rm x}_0=0$. Note, that the results for $N_{\rm tot}$ associated with a given choice of $T$, but different values of $M^2$ essentially fall on top of each other and cannot be discerned by eye. The longitudinal offset is fixed to ${\rm z}_0= M^2\mathfrak{z}_R=0.29\,\si{\meter}$; $\theta_{\rm s,x}=0$ and $\Delta\vartheta_{\rm s,x}=0$.}
	\end{figure}

	In Fig.~\ref{fig:Z0Y0}, we study the dependence of the angular shift $\Delta\vartheta_{\rm s,y}$, the divergence $\theta_{\rm s,y}$ and the signal photon number $N_{\rm tot}$ on the transverse offset ${\rm y}_0$ for a fixed longitudinal offset of ${\rm z}_0= M^2 \mathfrak{z}_R=\omega\mathfrak{w}_0^2/2=0.29\,\si{\meter}$.
	In accordance with the analytic estimate~\eqref{eq:deltathetas2}, the ratio $\Delta\vartheta_{\rm s,y}/M^2$ increases linearly with the transverse offset ${\rm y}_0$ for all considered choices of $T$ and $M^2$.
	While the curves extracted from the numerical calculation for both $M^2=1$ and $10$ essentially fall on top of the analytic estimate for $T=17\,{\rm fs}$, the curves for $T=220\,{\rm fs}$ deviate visibly from the analytic estimate.
	This behavior can again be attributed to the local nature of the approximation used to infer the estimate~\eqref{eq:deltathetas2}, which manifestly assumes the signal to originate from the vicinity of ${\rm z}={\rm z}_0$ only:
	as argued above, the smaller the value of $T$, the better this approximation for fixed other parameters.
	
	The reduction of the signal photon numbers in Fig.~\ref{fig:Z0Y0} with increasing values of ${\rm y}_0$ is consistent with a Gaussian decay of the form $\exp(-\#{\rm y}_0^2)$. This resembles the behavior inferred in Ref.~\cite{Karbstein:2018omb} resorting to an infinite Rayleigh range approximation for the probe beam. 
	Moreover, Fig.~\ref{fig:Z0Y0} confirms the prediction of Eq.~\eqref{eq:thetas2} that the radial divergence of the signal $\theta_{\rm s,y}$ in $\rm y$ direction essentially does not depend on ${\rm y}_0$ for $0\leq{\rm y}_0\leq\mathfrak{w}_0$.
	In fact, only for the case of $T=220\,{\rm fs}$ and $M^2=1$ a mild dependence on ${\rm y}_0$ is at all visible by eye.
	
	Next, we aim at assessing the perspectives of Setup B for quantum vacuum experiments.
	In this context, an important concept is the directional {\it discernibility} criterion comparing the number of signal photons emitted into a given solid angle interval ($\varphi,\vartheta$) in the far field with the number of probe photons to be detected in the same angular region. The latter constitute the background against which the signal photons are to be discriminated. Signal photons can be considered as discernible from the background if they fulfill the condition,
	\begin{equation}
		\frac{{\rm d}^2 N_p}{{\rm d}\varphi\,{\rm d}\!\cos\vartheta}\geq{\cal P}_p \frac{{\rm d}^2 \mathfrak{N}}{{\rm d}\varphi\,{\rm d}\!\cos\vartheta}\,,\label{eq:discern}
	\end{equation}
	where ${\cal P}_p>0$ denotes a polarization $p\in\{{\rm tot},\perp\}$ dependent detection efficiency \cite{Karbstein:2019oej}. In the present case we assume ${\cal P}_{\rm tot}=1$ for the signal attainable in a polarization insensitive measurement. On the other hand, for the measurement of polarization-flipped signal photons we identify ${\cal P}_\perp={\cal P}$, where $\cal P$ is the polarization purity of the employed polarimeter. In the latter case, the differential number of background photons registered in the perpendicularly polarized mode is given by ${\rm d}^2\mathfrak{N}_\perp={\cal P}\,{\rm d}^2\mathfrak{N}$. Here we use ${\cal P}=1.4 \times 10^{-11}$ \cite{Marx:2013xwa,Schulze:2018,Bernhardt:2020vxa,Schmitt:2020ttm}.
	
	Equation~\eqref{eq:thetas2} predicts the ratio of the signal and probe divergences to scale approximately as
	\begin{equation}
		\frac{\theta_{\rm s}}{\theta} \simeq \frac{1}{M^2}\sqrt{\frac{\{1+2(\frac{\mathfrak{w}_0}{w_0})^2[1+(\frac{{\rm z}_0}{\mathfrak{z}_R})^2]\}^2+M^4(\frac{{\rm z}_0}{\mathfrak{z}_{\rm R}})^2}{\{1+2(\frac{\mathfrak{w}_0}{w_0})^2[1+(\frac{{\rm z}_0}{\mathfrak{z}_R})^2]\}[1+(\frac{{\rm z}_0}{\mathfrak{z}_R})^2]}} \,. \label{eq:thetasbytheta}
	\end{equation}
	Hence, for the pump and probe waists characterizing Setup B and a probe beam quality factor of $M^2=1$, we can study the behavior of this ratio as a function of ${\rm z}_0/\mathfrak{z}_R$.
	While Eq.~\eqref{eq:thetasbytheta} varies between $\sqrt{19}\simeq4.36$ (maximum at ${\rm z}_0/\mathfrak{z}_R=0$) and $3\sqrt{2}\simeq4.24$ (value for ${\rm z}_0/\mathfrak{z}_R\to\infty$) for $M^2=1$, it just varies between $\simeq0.45$ (maximum at ${\rm z}_0/\mathfrak{z}_R\simeq0.84$) and $3\sqrt{2}/10\simeq0.42$ (value for ${\rm z}_0/\mathfrak{z}_R\to\infty$) for $M^2=10$.
	This suggests that for zero transverse offsets the best signal-to-background separation should be achievable with $M^2=1$ rather than $M^2=10$.
	Moreover, recall that only the choice of $M^2=1$ is compatible with the paraxial Helmholtz equation.
	Hence, even though the angular shift~\eqref{eq:deltathetas2} scales with a positive power of $M^2$, in the remainder of this work we exclusively provide results for a probe beam quality factor of $M^2=1$.
	
	In Tab.~\ref{tab:FutureBGPara} we showcase the parameters characterizing the angular signal photon distributions for Setup B and a selection of focal offsets: apart from collisions at zero offset, we consider a purely longitudinal focal offset ${\rm z}_0=\mathfrak{z}_R\simeq 0.29\,\si{m} $ by one probe Rayleigh range, as well as a combined transverse and longitudinal offset characterized by ${\rm z}_0= \mathfrak{z}_R$ and ${\rm y_0} = \mathfrak{w}_0\simeq 3.0\,\si{\micro m}$.
	\begin{table}
		\begin{tabular}{c|c|c||c|c|c|c|c|}
			$T$ [\si{\femto\second}] & ${\rm z}_0$ [\si{\meter}] & ${\rm y}_0$ [\si{\micro\meter}] & $\Delta\vartheta_{\rm s,y}$ [\si{\micro\radian}] & $\theta_{\rm s,x}$ [\si{\micro\radian}] & $\theta_{\rm s,y}$ [\si{\micro\radian}] & $N_{\rm tot}$ & $N_{\perp}$ \\
			\hline\hline
			\multirow{3}{*}{$17$} & $0$ & $0$ & $0.00$ & $41.53$ & $41.53$ & $6.68$ & $0.31$ \\
			& $0.29$ & $0$ & $0.00$ & $40.89$ & $40.89$ & $3.44$ & $0.16$ \\
			& $0.29$ & $3.0$ & $4.92$ & $40.89$ & $40.89$ & $1.31$ & $0.06$ \\
			\hline
			\multirow{3}{*}{$220$} & $0$ & $0$ & $0.00$ & $30.92$ & $30.92$ & $2.95$ & $0.14$ \\
			& $0.29$ & $0$ & $0.00$ & $29.20$ & $29.20$ & $1.56$ & $0.07$ \\
			& $0.29$ & $3.0$ & $4.53$ & $28.49$ & $28.49$ & $0.61$ & $0.03$
		\end{tabular}
		\caption{\label{tab:FutureBGPara} Parameters characterizing the signal photon distributions for Setup B with $M^2=1$. We consider two distinct probe pulse durations $T$, as well as different focal offsets; ${\rm x}_0=0$. The showcased results are inferred from a numerical evaluation of Eq.~\eqref{eq:d2N}. For all parameter sets featured in this table we find $\Delta\vartheta_{\rm s,x}\simeq0$.}
	\end{table}
	Depending on the focal offset we obtain about $N_{\rm tot}\simeq1\ldots7$ signal photons per shot. The polarization flipped signal $N_\perp$ is about a factor of $9/196\simeq0.046$ smaller; cf. Eq.~\eqref{eq:d2N}.
	In line with the above findings, finite focal offsets generically reduce the signal photon numbers $N_{\rm tot}$ and $N_\perp$. While the signal divergence decreases with ${\rm z_0}$, it remains practically unchanged when an additional transverse offset ${\rm y}_0\lesssim\mathfrak{w}_0$ is considered.
	
\begin{figure}
		\includegraphics[width=1.0\textwidth]{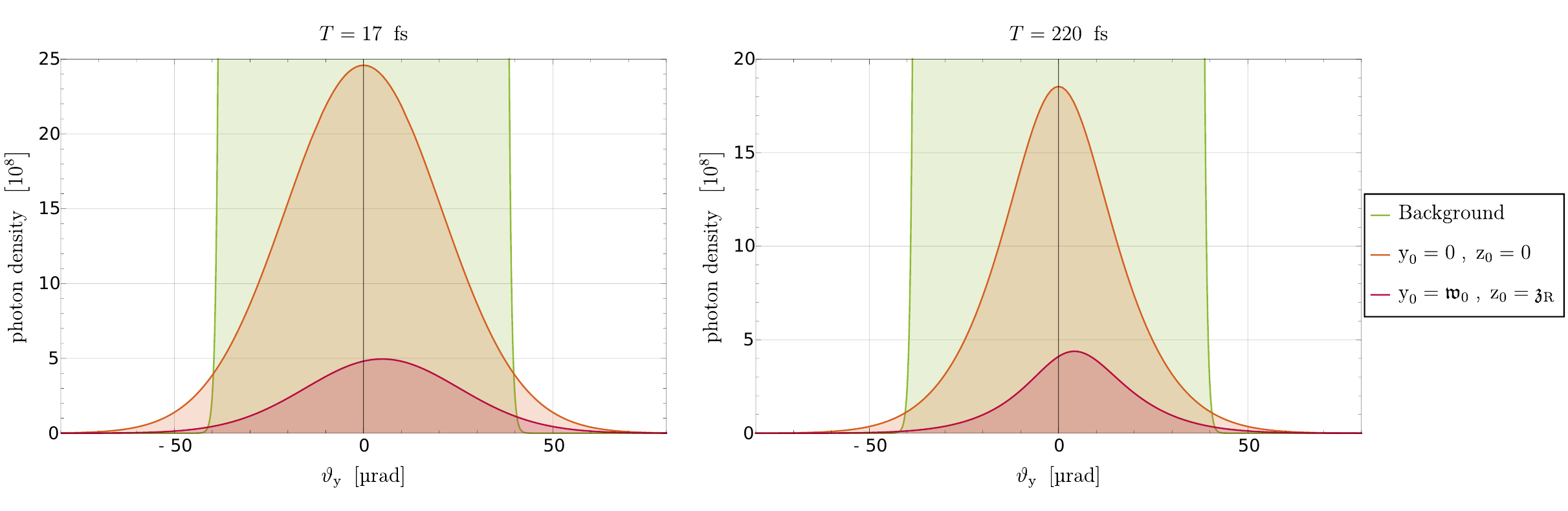}
		\caption{\label{fig:BackgroundB} Far-field angular distribution of the signal ${\rm d}^2N_{\rm tot}/({\rm d}\varphi\,{\rm d}\!\cos\vartheta)$ and the background ${\rm d}^2\mathfrak{N}/({\rm d}\varphi\,{\rm d}\!\cos\vartheta)$ photons for Setup B with $T=17\,{\rm fs}$ and $T=220\,{\rm fs}$ as a function of $\vartheta_{\rm y}$. Here, we show results for the signals attainable at zero  (middle curve) and finite (bottom) focal offsets. In all considered cases, the signal surpasses the background for $\vartheta_{\rm y}\gtrsim40 \, \si{\micro\radian}$.}
	\end{figure}

	In Fig.~\ref{fig:BackgroundB} we compare the far-field angular distributions of the signal photons $N_{\rm tot}$ and the background photons $\mathfrak{N}$ for Setup B as a function of $\vartheta_{\rm s,y}$. 
	Here, we depict results for a vanishing and a finite focal offset.
	In accordance with the above analytical scalings, for ${\rm z}_0=\mathfrak{z}_R$ and ${\rm y}_0=\mathfrak{w}_0$ the signal peak is notably shifted in the direction of the transverse offset, while the background remains invariant.
	In the vicinity of the forward beam axis of the probe beam the signal is vastly dominated by the background of the driving probe photons which traverse the interaction region with the pump pulse essentially unmodified.	
	However, from a certain value of the angle $\vartheta_{\rm y}$ onwards the signal surpasses the background.	
	For vanishing transverse focal offsets this happens solely because the divergence of the signal is larger than the divergence of the probe beam for $\mathfrak{w}({\rm z}_0)\gtrsim w_0$ \cite{Karbstein:2018omb}.
	On the other hand, for a combined longitudinal and transverse offset, the associated angular shift $\Delta\vartheta_{\rm s}$ results in an asymmetric photon distribution and thus potentially provides an additional means to improve the signal-to-background separation.
	At the same time, in accordance with the behavior observed in Fig.~\ref{fig:Z0Y0} and Tab.~\ref{tab:FutureBGPara}, a transverse offset generically comes with a reduction of the signal photon number.
	\begin{figure}
		\includegraphics[width=1.0\textwidth]{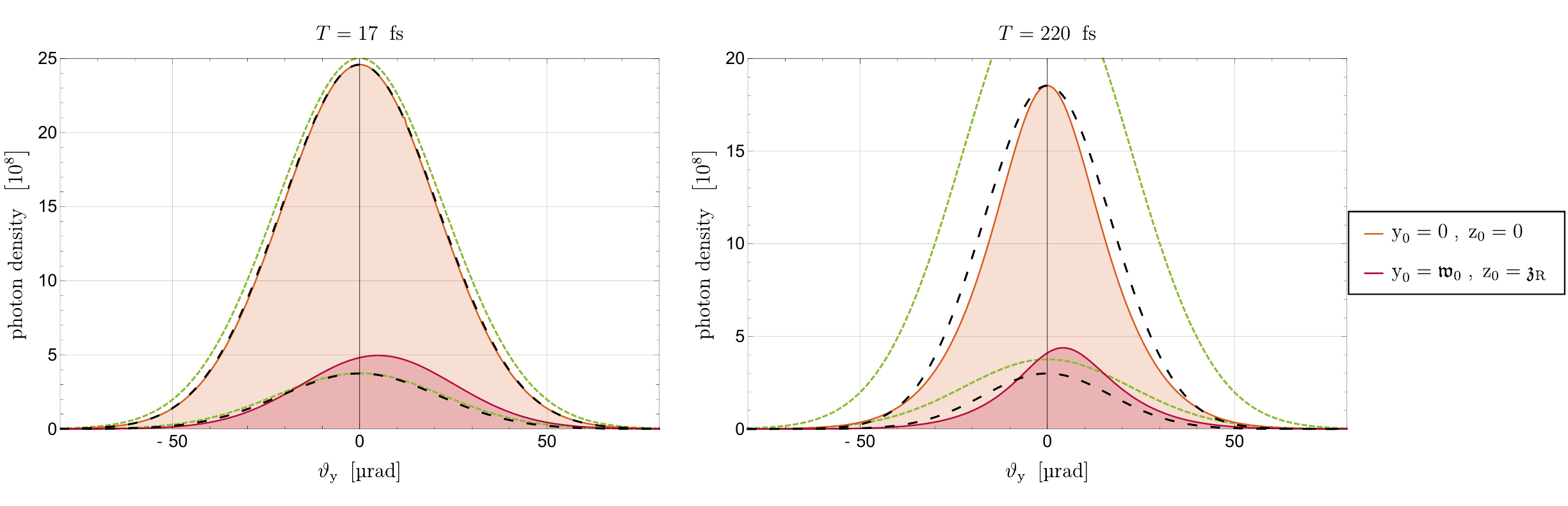}
		\caption{\label{fig:InfAppFin}  Far-field angular distribution ${\rm d}^2N_{\rm tot}/({\rm d}\varphi\,{\rm d}\!\cos\vartheta)$ of the signal photons for Setup B with $T=17\,{\rm fs}$ and $T=220\,{\rm fs}$ as a function of $\vartheta_{\rm y}$. Here we confront the full numerical results with the outcomes of analytical approximations devised for focal offsets fulfilling ${\rm z}_0\ll\mathfrak{z}_R$, namely the infinite Rayleigh range approximation (green dashed) and the analytical approximation of Ref.~\cite{Mosman:2021vua} (black dashed). The upper set of curves is for zero offset and the lower one for ${\rm y}_0=\mathfrak{w}_0$, ${\rm z}_0=\mathfrak{z}_R$.}
	\end{figure}
	\begin{table}
		\begin{tabular}{c|c|c||c|c|c|}
			\multicolumn{3}{c||}{} & \multicolumn{3}{c|}{$N_{\rm tot}$} \\
			$T$ [\si{\femto\second}] & ${\rm z}_0$ [\si{\meter}] & ${\rm y}_0$ [\si{\micro\meter}] & IRRA & Ref.~\cite{Mosman:2021vua} & Numerical \\
			\hline\hline
			& $0$ & $0$ & $7.75$ & $6.76$ & $6.68$ \\
			$17$ & $0.29$ & $0$ & $3.98$ & $3.48$ & $3.44$ \\
			& $0.29$ & $3.0$ & $1.50$ & $1.32$ & $1.31$ \\
			\hline
			& $0$ & $0$ & $7.74$ & $3.38$ & $2.95$ \\
			$220$ & $0.29$ & $0$ & $3.98$ & $1.84$ & $1.56$ \\
			& $0.29$ & $3.0$ & $1.50$ & $0.71$ & $0.61$
		\end{tabular}
		\caption{\label{tab:Fig7Tab} Integrated numbers of signal photons $N_{\rm tot}$ associated with the distributions shown in Fig.~\ref{fig:InfAppFin}. As to be expected, the infinite Rayleigh range approximation (IRRA) overestimates the signal. At the same time, the values predicted by the analytical approximation~\cite{Mosman:2021vua} are much closer to the numerical results.}
	\end{table}
	For finite transverse focal offsets (${\rm z}_0$, ${\rm y}_0$) the signal photon distribution exhibits an asymmetry with respect to the forward beam axis of the probe beam.
	Hence, for a given choice of $p\in\{{\rm tot},\perp\}$ the discernibility criterion~\eqref{eq:discern} is generically met for two different $p$-dependent values $\vartheta_{{\rm y},{\rm L}/{\rm R}}$ of the angle $\vartheta_{\rm y}$ on the left (L) and on the right (R) side of $\vartheta_{\rm y}=0$.
	
	In an attempt to further understand the imprint of probe wave-front curvature on the signal photon distributions, we compare the numerical results with the outcomes of analytical approximations which manifestly neglect this effect: these are the infinite Rayleigh range approximation introduced in Sec.~\ref{sec:formalism}, and the more advanced analytical approximation put forward in Ref.~\cite{Mosman:2021vua}; in the latter we identify $w_{\rm x}=\mathfrak{w}({\rm z}_0)$. As opposed to the infinite Rayleigh range approximation, which is based on the formal limit of $\{\mathfrak{z}_R,{\rm z}_R\}\to\infty$, the latter one only utilizes $\mathfrak{z}_R\to\infty$, but explicitly accounts for the finite Rayleigh range ${\rm z}_R$ of the pump. We emphasize that discrepancies between these approximate results and the numerical calculation are to be expected because the regime of applicability of the former is a priori restricted to longitudinal offsets ${\rm z}_0\ll\mathfrak{z}_R$, where probe wave-front curvature effects can be safely neglected.
	
	While the curves for both approximations are close to each other for $T=17\,{\rm fs}$, clear discrepancies are visible for $T=220\,{\rm fs}$. This behavior can be explained by the fact that in the first case we have ${\rm z}_R\sim T$, while in the second one we have $T\gg{\rm z}_R$, such that -- aside from the wavefront curvature effects, which are manifestly neglected from the outset -- the criterion for the infinite Rayleigh range approximation to allow for trustworthy results, $\{T,\tau\}\ll\{\mathfrak{z}_R,{\rm z}_R\}$, is severely violated. In line with that, for $T=220\,{\rm fs}$ the results of the more advanced analytical approximation of Ref.~\cite{Mosman:2021vua} are substantially closer to the full numerical results. In Tab.~\ref{tab:Fig7Tab} we provide the associated signal photon numbers.
	It is worth noting that the infinite Rayleigh range approximation is essentially insensitive to variations in $T$, resulting in relatively large discrepancies to the numerical results especially for the larger value of $T$. 
	On the other hand, the quite good agreement of the signal photon numbers derived on the basis of the analytical approximation~\cite{Mosman:2021vua} and the numerical results hints at the fact that discrepancies between the outcomes of this approximation and the full results can be mainly attributed to wave-front curvature effects: while the wave-front curvature of the probe beam in the interaction region with the pump laser pulse can certainly impact the far-field directional distribution of the signal photons, it should not substantially alter the associated signal photon numbers.
	
	In fact, the rather good agreement of the predicted signal photon numbers suggests that a combination of the analytical approximation~\cite{Mosman:2021vua} with the estimates~\eqref{eq:thetas} and \eqref{eq:deltathetas} derived in the present work can be straightforwardly combined to obtain a rough analytical estimate of the far-field angular distribution of the signal photons, which phenomenologically accounts for probe wave-front curvature effects in laser pulse collisions at large focal offsets. To this end, the far-field angular distribution of the signal is to be parameterized by 
	\begin{equation}
		\frac{{\rm d}^2N_p}{{\rm d}\varphi\,{\rm d}\!\cos\vartheta}=n_p\,{\rm e}^{-2(\frac{\vartheta-\Delta\vartheta_{\rm s}(\varphi)}{\theta_{\rm s}(\varphi)})^2}\,,\ \text{with}\ \int_0^{2\pi}{\rm d}\varphi\int_0^\infty{\rm d}\vartheta\,\vartheta\,\frac{{\rm d}^2N_p}{{\rm d}\varphi\,{\rm d}\!\cos\vartheta}\simeq N_p|_\text{\cite{Mosman:2021vua}}\,,
	\end{equation}
	where the latter condition fixes the amplitude $n_p$.
	
	Finally, in Tabs.~\ref{tab:PartialNFull} and \ref{tab:PartialN} we analyze the numbers of discernible signal photons $N_{p,{\rm L}/{\rm R}}$ per shot emitted to the left of $\vartheta_{{\rm y},{\rm L}}$ and to the right of $\vartheta_{{\rm y},{\rm R}}$ but arbitrary values of $\vartheta_{\rm x}$, respectively. The corresponding numbers of background photons are $\mathfrak{N}_{p,{\rm L}/{\rm R}}$.
	For ${\rm y}_0=0$ we clearly have $|\vartheta_{\rm y, L}|=|\vartheta_{\rm y,R}|$. As a measure of the asymmetry of the far-field angular distribution of the photons of frequency $\simeq\omega$ we introduce the ratio $( N_{p,{\rm L}} + \mathfrak{N}_{p,{\rm L}})/(N_{p,{\rm R}} + \mathfrak{N}_{p,{\rm R}})$ between the number of $p$-polarized (signal+background) photons registered at angles to the left/right of $\vartheta_{\rm y, L/R}$.
	\begin{table}
		\begin{tabular}{c|c|c||c c|c c|c|}
			$T$ [\si{\femto\second}] & ${\rm z}_0$ [\si{\meter}] & ${\rm y}_0$ [\si{\micro\meter}] & $|\vartheta_{\rm y,L}|$ [\si{\micro\radian}] & $N_{{\rm tot},{\rm L}}$ & $|\vartheta_{\rm y,R}|$ [\si{\micro\radian}] & $N_{{\rm tot},{\rm R}}$ & $\frac { N_{\rm tot,R} + \mathfrak{N}_{\rm R} }{ N_{\rm tot,L} + \mathfrak{N}_{\rm L} }$ \\
			\hline\hline
			& $0$ & $0$ & $39.69$ & $0.19$ & $39.69$ & $0.19$ & $1$ \\
			$17$ & $0.29$ & $0$ & $40.17$ & $0.089$ & $40.17$ & $0.089$ & $1.00$ \\
			& $0.29$ & $3.0$ & $41.18$ & $0.030$ & $40.53$ & $0.032$ & $0.92$ \\
			\hline
			& $0$ & $0$ & $40.53$ & $0.044$ & $40.53$ & $0.044$ & $1$ \\
			$220$ & $0.29$ & $0$ & $41.01$ & $0.020$ & $41.01$ & $0.020$ & $1.00$ \\
			& $0.29$ & $3.0$ & $42.06$ & $0.0067$ & $41.31$ & $0.0073$ & $0.89$
		\end{tabular}
		\caption{\label{tab:PartialNFull} Discernible numbers of signal photons $N_{\rm tot,L/R}$ scattered to the left/right of $\vartheta_{\rm y,L/R}$, respectively; $M^2=1$ and ${\rm x}_0=0$. In the last column we provide the ratio of the numbers of (signal+background) photons measured to the left/right of $\vartheta_{\rm y,L/R}$; a ratio different from unity signalizes an asymmetry in the far-field distribution of the photons of frequency $\simeq\omega$.}
	\end{table}
	\begin{table}
		\begin{tabular}{c|c|c||c c|c c|c|}
			$T$ [\si{\femto\second}] & ${\rm z}_0$ [\si{\meter}] & ${\rm y}_0$ [\si{\micro\meter}] & $|\vartheta_{\rm y,L}|$ [\si{\micro\radian}] & $N_{\perp,{\rm L}}$ & $|\vartheta_{\rm y, R}|$ [\si{\micro\radian}] & $N_{\perp,{\rm R}}$ & $\frac { N_{\perp,{\rm L}} + \mathfrak{N}_{\perp,{\rm L}} }{ N_{\perp,{\rm R}} + \mathfrak{N}_{\perp,{\rm R}} }$ \\
			\hline\hline
			& $0$ & $0$ & $19.03$ & $0.056$ & $19.03$ & $0.056$ & $1$ \\
			$17$ & $0.29$ & $0$ & $19.95$ & $0.026$ & $19.95$ & $0.026$ & $1.00$ \\
			& $0.29$ & $3.0$ & $21.61$ & $0.0088$ & $20.96$ & $0.0092$ & $0.94$ \\
			\hline
			& $0$ & $0$ & $20.08$ & $0.017$ & $20.08$ & $0.017$ & $1$ \\
			$220$ & $0.29$ & $0$ & $20.99$ & $0.0079$ & $20.99$ & $0.0079$ & $1.00$ \\
			& $0.29$ & $3.0$ & $22.82$ & $0.0025$ & $21.77$ & $0.0028$ & $0.88$
		\end{tabular}
		\caption{\label{tab:PartialN} Discernible numbers of signal photons $N_{\perp,{\rm L/R}}$ scattered to the left/right of $\vartheta_{\rm y,L/R}$, respectively; $M^2=1$ and ${\rm x}_0=0$. In the last column we provide the ratio of the numbers of $\perp$-polarized photons registered to the right and left of $\vartheta_{\rm y,L/R}$.}
	\end{table}
	
	As the peak value of the background is larger for $N_{\rm tot}$ than for $N_\perp$, for a given offset the moduli of the discernibility angles $|\vartheta_{\rm y,L/R}|$ are generically larger than the former. Nevertheless, the discernible numbers of signal photons for $p={\rm tot}$ surpass those for $p=\perp$ which can be detected using high-precision polarimetry: the polarization-insensitive signals highlighted in Tab.~\ref{tab:PartialNFull} are about a factor of two to three larger than the $\perp$-polarized ones in Tab.~\ref{tab:PartialN}. For the larger probe pulse duration $T$ the asymmetry of the far-field photon distribution encountered for the combined longitudinal and transverse focal offset is somewhat larger than that for the smaller one. Presuming a repetition rate of $1\,{\rm Hz}$, for the combined longitudinal and transverse offset ${\rm z}_0=0.29\,\si{\meter}$, ${\rm y}_0=3.0\,\si{\micro \meter}$ and a pulse duration of $T=220\,\si{\femto \second}$ resulting in the largest asymmetry for the data sets studied here, we obtain $N_{\rm tot,L}\simeq24/{\rm h}$ ($N_{\perp,{\rm L}}\simeq9/{\rm h}$) and $N_{\rm tot,R}\simeq26/{\rm h}$ ($N_{\perp,{\rm R}}\simeq10 /{\rm h} $). The discernible signal photon numbers attainable at zero offset are much larger, namely $N_{\rm tot, L}= N_{\rm tot, R}\simeq158 /{\rm h} $ ($N_{\perp,{\rm L}}=N_{\perp,{\rm R}}\simeq61 /{\rm h} $).

	\section{Conclusions}\label{sec:conclusions} %%%%%%%%%%%%%%%%%%%%%%
	
	In the present article we studied the non-linear QED signature of vacuum diffraction in the head-on collision of an x-ray probe with a high-intensity laser pump at large spatio-temporal offsets between the beam foci. Both probe and pump fields are described as paraxial Gaussian beams supplemented with a temporal pulse envelope. In our explicit examples the beam radius of the probe in the interaction region with the pump is always larger than the beam radius of the pump.
	In accordance with naive expectations, for longitudinal offsets of the order of the probe Rayleigh range and transverse offsets of the order of the probe waist wavefront curvature effects of the probe can no longer be neglected and leave an imprint on the far-field directional distribution of the signal photons. Most prominently, this can result in a finite angular shift of the main emission direction of the signal away from the forward beam-axis of the probe.
	
	The present work is motivated by the recent work~\cite{Seino}, which predicted the emergence of this particular feature on the basis of a phenomenological ad hoc model involving several simplifying assumptions. As opposed to Ref.~\cite{Seino}, our study analyzes the the emergence of this feature self-consistently on the basis of a paraxial beam model.
	The far-field angular distribution of the signal photons is studied in detail, and compact analytical scalings for the far-field radial divergence of the signal and its angular shift relatively to the forward beam-axis of the probe laser are derived.
	
	Special attention is put on discernible signals in prospective discovery experiments of QED vacuum nonlinearity.
	To this end, we consider both polarization-insensitive measurements and the detection of polarization-flipped signal photons utilizing high-definition polarimetry. Our results allow to assess the impact of large focal offsets on the attainable 
	signals.
	Hence, they are relevant for the analysis of experimental scenarios suffering from large shot-to-shot fluctuations, which result in non-optimal collisions.
	We put special attention on the asymmetry imprinted on the far-field angluar distribution of the signal by utilizing both finite longitudinal and transverse focal offsets, and characterize it in terms of the ratio of the photons measured to the left/right of the angles $\vartheta_{\rm y, L/R}$.
	
	It would be very interesting to repeat the present analysis with probe beams featuring different transverse profiles in the interaction region with the high-intensity laser pulse, such as top-head or flattened-Gaussian beams \cite{Karbstein:2020gzg}: while  the transverse decay of the Gaussian probe inevitably comes with a reduction of the photons available for probing at finite transverse offsets, a flattened beam would not suffer from this loss. This could provide a means to increase the discernible signals.

	\acknowledgments %%%%%%%%%%%%%%%%%%%%%%%%%%%%%%%%%%%%%%%%%%%%%%%%%%
	This work has been funded by the Deutsche Forschungsgemeinschaft (DFG) under Grant No. 416607684 within the Research Unit FOR2783/1. The authors are grateful to Yudai Seino and Toshiaki Inada for helpful correspondence and fruitful discussions.

\end{document}